\documentclass[journal]{amsart}
\usepackage{graphicx,tikz} 
\usepackage{amssymb}
\usepackage{amsmath}
\usepackage{enumerate}
\begin{document}

\title{Time-Frequency Filtering Meets Graph Clustering}

\author{Marcelo A. Colominas, Stefan Steinerberger, and Hau-Tieng Wu}

\maketitle

\begin{abstract}
We show that the problem of identifying different signal components from a time-frequency representation can be equivalently phrased as a graph clustering problem: given a graph $G=(V,E)$ one aims to identify `clusters', subgraphs that are strongly connected and have relatively few connections between them. The graph clustering problem is well studied, we show how these ideas can suggest (many) new ways to identify signal components. Numerical experiments illustrate the ideas.
\end{abstract}

\section{Introduction}
Linear time-frequency (TF) representations are a versatile and powerful tool to analyze multicomponent nonstationary signals. Among them, the Short-Time Fourier Transform (STFT) is perhaps the most natural one, since it amounts to locally analyzing the signal frequency content by windowing the signal with a fast-decaying function \cite{Cohen1995,flandrin2018explorations}. When signals have a finite number of components (such as AM-FM oscillations and/or impulse-like transients with sufficient separation), the TF representation (TFR) determined by STFT presents a special structure: signal information is confined to a few ``ribbons'' organized around the instantaneous frequency (IF) of each component. Synchrosqueezing \cite{Daubechies1996,Daubechies2011,Oberlin2015} the STFT (either vertically or horizontally) amounts to reassign the coefficients so these ribbons are ``narrower'': the signal information is concentrated, while the representation remains invertible. Filtering this type of signals in order to reconstruct (and denoise) different components is achieved by localizing these TF regions (the mentioned ribbons), and inverting the representation using only the coefficients ``ribbon-wise''. This can be seen as a {\em masking problem} on the TF plane.

A common strategy for filtering is the identification of {\em curves} in TFR called \emph{ridges}, which are estimations of the IFs of the signal components \cite{Carmona1997,Carmona1999,liu2024analyzing}. After this, a ribbon is defined around the ridge, which can have either a constant or a variable width \cite{colominas2020fully}. A deflationary scheme (so-called peeling algorithm) is applied, so each TF region is extracted \emph{sequentially}, one after another. This procedure is prone to error accumulation, since each TF region is ``deleted'' after being extracted. We propose here a graph theory approach, which is a flexible approach with three main advantages: i) it does \textit{not} assume the ridge duration to be equal to that of the signal (an assumption often made in ridge extraction), making the identification of transient components easier; ii) its non-parametric approach offers a rotational invariance in the TF plane, being equally applicable to AM-FM oscillations (horizontal TF signatures) and impulse-like transients (which have a vertical TF signature); and iii) it does not work sequentially, meaning all the components are extracted \emph{simultaneously}, thus avoiding the error accumulation problem of the peeling algorithm.

The paper is organized as follows. Sec. \ref{sec:TFR} recalls some basics concepts of TF representations, while Sec. \ref{sec:Graph} introduces our main contribution. Numerical examples are shown in \ref{sec:Examples}, while Sec. \ref{sec:Conclusions} concludes the paper.

\section{Time-Frequency Representations}\label{sec:TFR}
\subsection{Time-Frequency Basics}
We will work in a discrete-time setting, where $x[n] = x(n \Delta t)$ is a sampled version of a signal $x(t)$ (here $\Delta t$ is the sampling period). In this setting, the STFT reads
$$F_x^g[m,n] = \sum_u x[u] g[u-n] e^{-i2\pi m [u-n]/M} $$
\noindent for $n = 0,1,\dots, N-1$ and $m = 0, 1, \dots,M-1$, and where the representation has $M$ frequency bins.
A synchrosqueezed version (where the information is more concentrated around the IF \cite{Daubechies2011}) can be obtained by
$$S_x^g[m,n] = \sum_v  F_x^g[v,n] \delta[m-\omega_x[v,n]]$$
\noindent where $\delta[\cdot]$ stands for the Kronecker delta, and $\omega_x[m,n]$ is the so-called {\em reassignment} operator that describes how to vertically reassign the coefficients \cite{Oberlin2015,Wu2011adaptive,pham2017high}.

\subsection{Time-Frequency Filtering}

Time-Frequency filtering exploits the invertibility  of linear time-frequency representations. For a representation $R_x$ (which can be either $F_x^g$ of $S_x^g$), a \emph{filtered} version of the input signal can be achieved as

$$\tilde{x}[n] = \frac{1}{g[0]} \sum_v R_x[v,n] \mathcal{M}[v,n]$$

\noindent where $\mathcal{M}$ is a time-frequency \emph{binary mask}. The filtered version depends on the mask, and it can be a denoised version of the input signal, or a specific component that one has interest in. This is of special interest when working with \emph{wave-shape functions} \cite{Wu2013} since one needs to obtain both amplitude and phase of the fundamental component.

\subsection{Existing Approaches}

Existing approaches can be classified into two main categories: parametric and non-parametric methods.
Among the parametric ones, the most widely used method is the one based on ridge detection, followed by the definition of a ribbon around the ridge. Variations can be found in the way the ridge is detected \cite{meignen2017demodulation,colominas2019time} and the ribbon is defined \cite{colominas2020fully}. This approach faces difficulties when dealing with ridges that cannot be parameterized as explicit functions of time (such as those components with vertical or circular TF signature).
Among the non-parametric approaches we can mention the one based on zeros of the STFT \cite{flandrin2017spectrogram}. Variations can be found on how to use such zeros for TF filtering, with examples using Delaunay triangulations \cite{flandrin2017spectrogram} and so-called ``holes'' \cite{bardenet2020zeros}. Examples where the zeros are further classified into different classes improves the performances \cite{miramont2024unsupervised}.

\section{A Graph Clustering Approach}\label{sec:Graph}
\subsection{Description.}
We now describe a (large) family of algorithms that offer a way to solve the problem by first turning it into a well-studied problem in graph theory. We assume that the time-frequency data is given to us in a discrete data format, as an image that is encoded as a matrix $A \in \mathbb{R}^{m \times n}$. This means that $A$ is the modulus of the time-frequency representation.

\begin{figure}[h!]
    \begin{tikzpicture}
 \node at (0,0) {\includegraphics[width=0.22\textwidth]{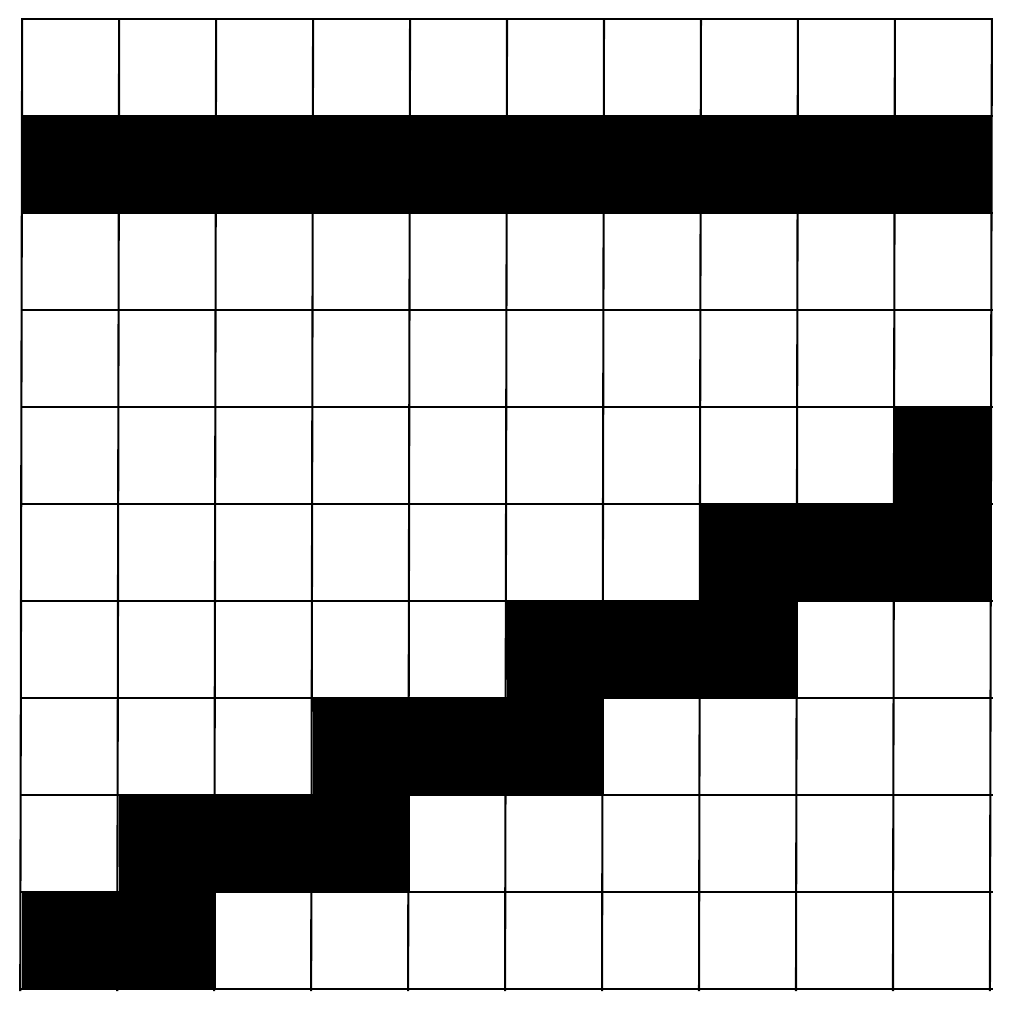}};
  \node at (4.3,0) {\includegraphics[width=0.22\textwidth]{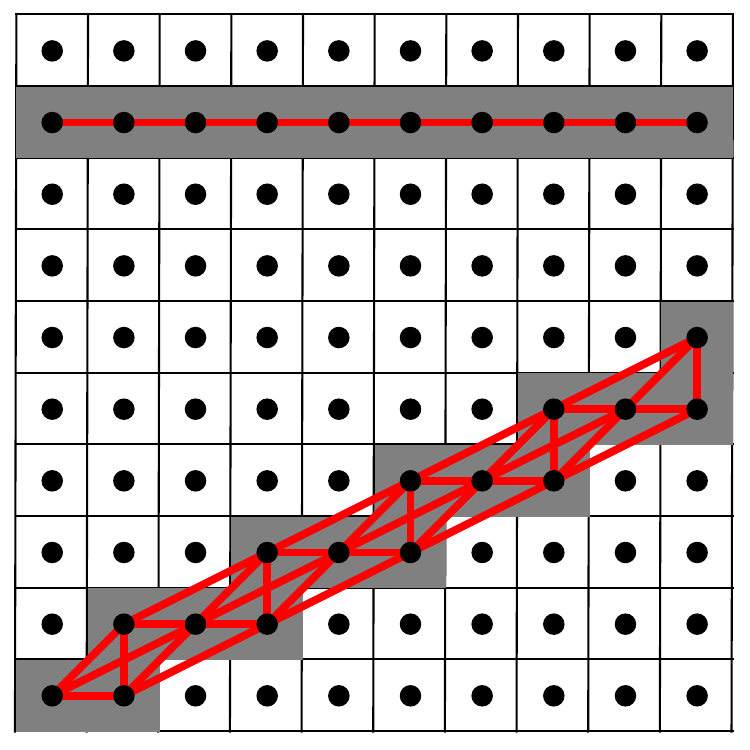}};
\end{tikzpicture}
\caption{Left: a (very simple) discrete time-frequency representation. Right: pixels are turned into
vertices of a graph. We connect two vertices with an edge if the corresponding two pixels are `close' in
the time-frequency plane and have large TF energy.}
\end{figure}

A basic template for an algorithm is now as follows (see also Fig. 1 for a visualization of what this looks like).

\textbf{Algorithm Template.}
\begin{enumerate}
    \item Generate a graph $G=(V,E)$ with vertex set $V = \left\{1,2,\dots, m\cdot n\right\}$ corresponding to the pixels.
    \item Connect two vertices $v,w$ by an edge $(v,w) \in E$ if
    \begin{enumerate}
        \item the pixels are `close' in the TF plane
        \item and the time-frequency entry in both pixels is `large'
    \end{enumerate}
    \item (\textit{Basic.}) Compute each connected component of the graph $G$, sorting them by their number of edges. Each of these corresponds to a separate TF domain. (\textit{Advanced.}) Run a graph clustering algorithm to identify components.
\end{enumerate}

A number of different algorithms can be obtained by specifying what precisely one means by `close' in the TF plane, what it means for both entries to be large and whether and what type of graph clustering algorithm is employed. The advantage of this method is that the graph clustering problem is incredibly well studied (see, for example, \cite{chao2021survey, ding2024survey, wang2023overview}). If we consider the toy problem in Fig. 1, then the separation problem is trivial: the TF energy is $\left\{0,1\right\}$ and the two components are clearly separated  The strength of the proposed class of algorithms comes from the fact that the graph clustering methods are very robust and can handle complicated graphs. In some cases, these methods are close to the information-theoretic limit \cite{abbe2015exact, abbe2018community}.

\begin{center}
\begin{figure}
\includegraphics[width=\columnwidth]{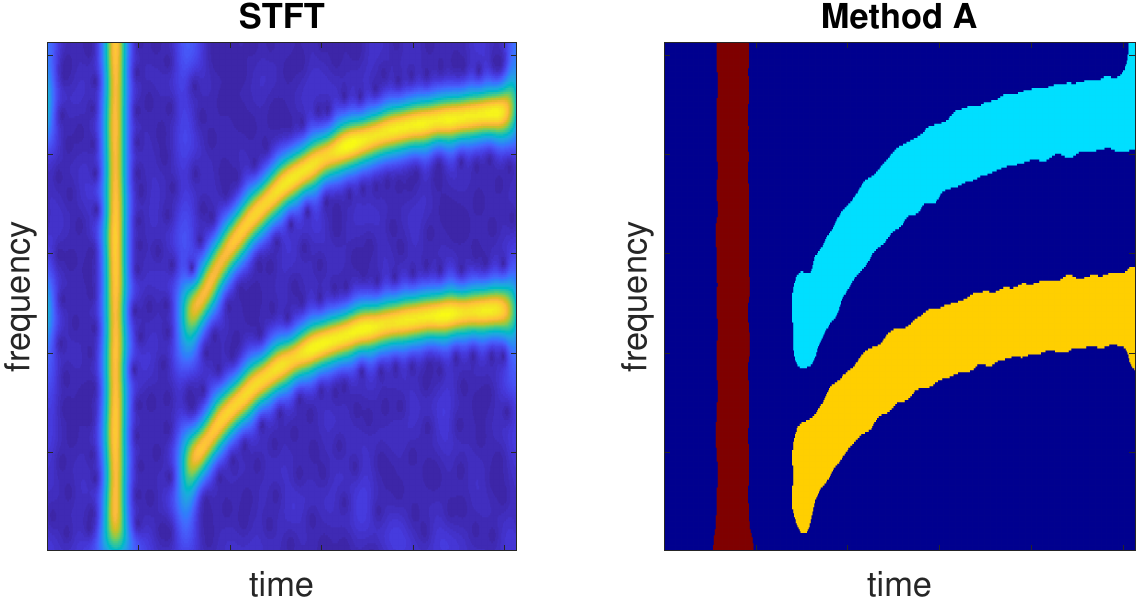}\caption{Impulse and chirps. Left: modulus of the STFT with an SNR of 20 dB. Right: selected masks. } \label{fig:Impulse}
\end{figure}
\end{center}

To make matters concrete, we propose \textbf{Method A}. This method defines two pixels $(i,j)$ and $(k,\ell)$ to be close when their distance is below some threshold in $\ell^1$
  $$ |i -k| + |j - \ell| \leq r,$$
and it says that these two pixels both carry large TF energy if
$$    | A_{ij}| \cdot |A_{k \ell}| \geq \tau$$
for some threshold $\tau$. The threshold $\tau$ can be defined as related to the noise level of the input signal. When the noise is Gaussian white, its standard deviation can be estimated as
\begin{equation} \label{eq:gamma}
\gamma = \sqrt{2} \cdot \mbox{median}(|\mbox{Re}(F_x^g)|)/0.6745,
\end{equation}
where $\mbox{Re}(\cdot)$ stands for the real part, and a threshold of $3\gamma$ is set to avoid 99\% of the noise \cite{Donoho1994}. When the noise is non-Gaussian, the Gaussianity of TFR as a random field is guaranteed so that this estimator can be considered \cite{wu2025uncertainty}. In such a way, we define the threshold as $\tau = 9\gamma^2$. We will refer to this combination ($A = |F_x^g|$ and $\tau = 9\gamma^2$) as \textbf{Method A}. An example of our algorithm applied to a multicomponent nonstationary signal can be observed in Fig. \ref{fig:Impulse} where we analyzed a signal composed of an impulse and two increasing exponential chirps.

\subsection{Variations of the method}
This type of algorithm has a lot of freedom in how things are set up. There is a choice in the notion of pixels being adjacent given by the parameter $r \in \mathbb{N}$. If $r$ is small, then ridges have to continuously have large time-frequency energy, if $r$ is slightly larger, then small gaps (possibly caused by noise) are allowed. One could replace the $\ell^1-$norm by the $\ell^2-$norm (especially when $r$ gets larger) but one could also replace it by $\ell^p$ with $1 \leq p \leq \infty$. The second ingredient is to connect pixels that both have large time-frequency energy, a natural choice being $ |A_{ij}| \cdot |A_{k \ell}| \geq \tau$. However, we note that
 \begin{equation}\label{eq:minimum}
 \min \left\{  |A_{ij}|,|A_{k \ell}|  \right\} \geq \tau
 \end{equation}
could also be a good choice and many other reasonable choices are conceivable. There is a wide variety of graph clustering algorithms that could be employed -- we will not do this here and show that already the most basic version of the method gives excellent results. For the sake of illustrating the breadth of possibilities, we consider four other methods.

\textbf{Method B.} As a second version for our algorithm, we will use $A = |F_x^g|$, and the connection policy (2), with $\tau =3 \gamma$. Notice that this would be equivalent to \emph{binarize} $|F_x^g|$ using \eqref{eq:gamma} for the threshold. 

\textbf{Method C.} The third variation will be one using $A = |S_x^g|$ (second-order SST) and connect when $ |A_{ij}| \cdot |A_{k \ell}| \geq 9\gamma^2/\sigma^2$. Indeed, when synchrosqueezing the STFT, one ends up with the amplitude of the new ridges being those of the original ridges divided by $\sigma$ \cite{Daubechies2011} (in the case of our normalization).

\textbf{Method D.} The fourth variation will use $A = |S_x^g|$ and connect whenever
$ \min \left\{  |A_{ij}|,|A_{k \ell}|  \right\} \geq 3\gamma/\sigma$. This would be equivalent to binarize $|S_x^g|$ using $3\gamma/\sigma$ as the threshold.

\textbf{Method E.} Finally, we will include a recent method described in \cite{zhou2025stft}. Originally conceived as an algorithm unrelated to graph theory, it can be considered as fitting into our framework. In this case, the matrix $A$ corresponds to the binarization of the smoothed STFT modulus. The authors perform this smoothing because in moderate to high noise situations, and when one has moderate to strong frequency modulation, the component TF domain can be ``broken''. So with this smoothing they try to recover this TF domain. For the binarization, they use an \emph{ad hoc} threshold which produces reasonably good results.

\begin{center}
\begin{figure*}
\includegraphics[width=\textwidth]{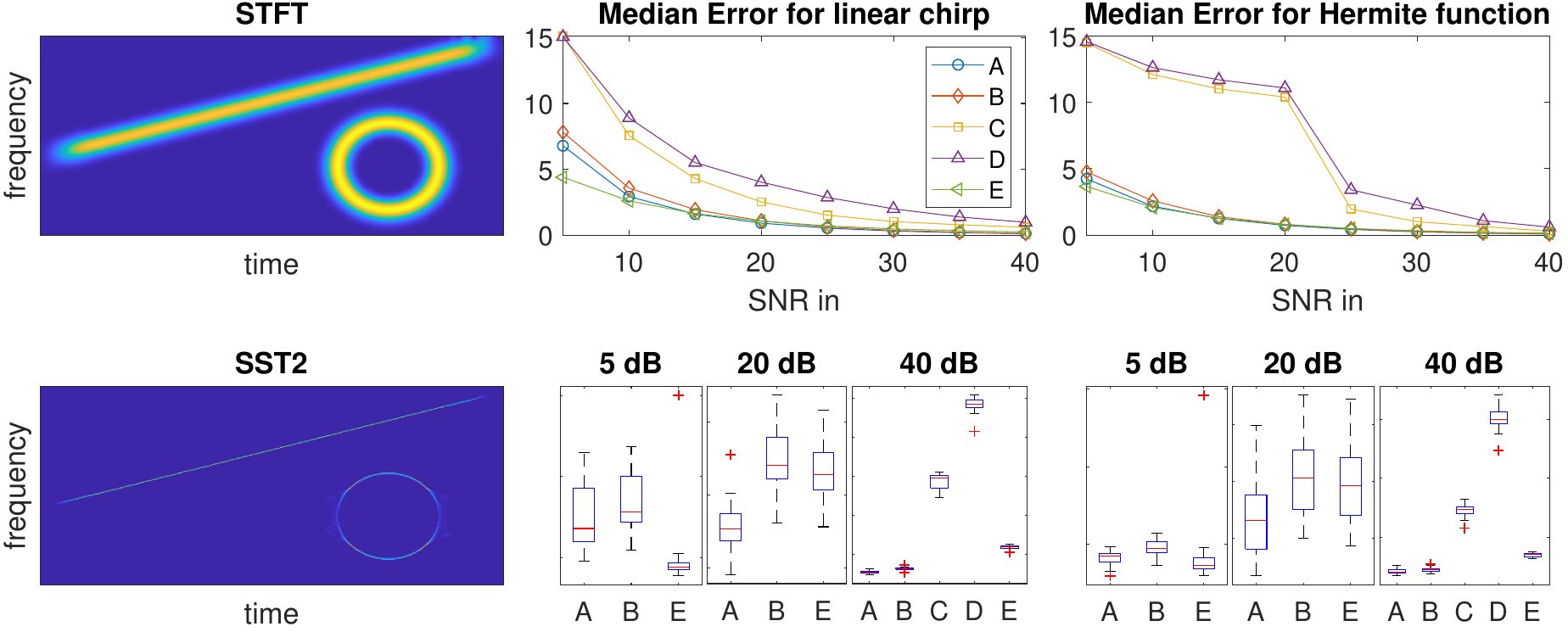}
    \caption{Linear chirp and Hermite function. Here we show the median over 30 realizations. Top left: modulus of the STFT for an SNR of 40 dB. Top middle: median errors for the linear chirp estimation. Top right: median errors for the Hermite function estimation. Bottom left: modulus of SST2 for an SNR of 40 dB. Bottom middle: boxplots for the errors when estimating the linear chirp. Bottom right: boxplots for the errors when estimating the Hermite function.}\label{fig:Hermite}
    \end{figure*}
\end{center}

\section{Numerical Examples}\label{sec:Examples}
\subsection{Simulated Signals}
\subsubsection{Linear chirp and Hermite function} We start by testing the different methods on a two-component signal, with a linear chirp and a Hermite function of order 20. Both components have the same energy. Results can be appreciated on Fig. \ref{fig:Hermite}.
We can see that in general, all combinations produce acceptable results. Under a high-noise situation (SNR of 5 dB), \textbf{Method E} seems to be superior, with two outliers however. These outliers happen when the method is not able to separate the two components and treat the signal as a single component. For moderate levels of noise (SNR of 20 dB), \textbf{Method A} is superior, and the same happens for a low level of noise (SNR of 40 dB). As for the Hermite function, we can see that the two methods based on the synchrosqueezed STFT (\textbf{Methods C and D}) underperform for high to moderate levels of noise. This is caused by the difficulties posed to SST by the almost vertical parts of the Hermite function TF signature.
\subsubsection{Sinusoidal and linear chirps} 
\begin{center}
\begin{figure*}
\includegraphics[width=\textwidth]{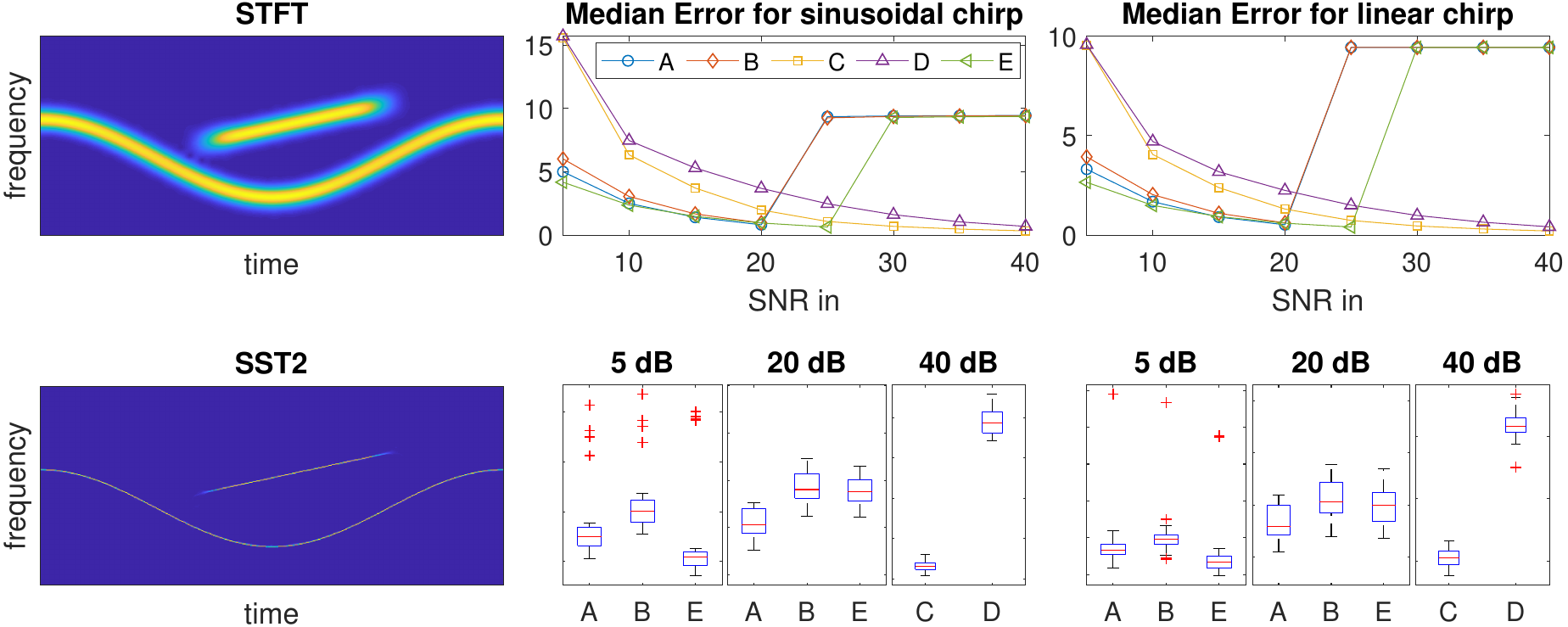}
    \caption{Sinusoidal and linear chirps. Here we show the median over 30 realizations. Top left: modulus of the STFT for an SNR of 40 dB. Top middle: median errors for the sinusoidal chirp estimation. Top right: median errors for the linear chirp estimation. Bottom left: modulus of SST2 for an SNR of 40 dB. Bottom middle: boxplots for the errors when estimating the sinusoidal chirp. Bottom right: boxplots for the errors when estimating the linear chirp.}\label{fig:Sinusoidal}
    \end{figure*}
\end{center}

As a second simulated example, we present the results on a signal composed of a sinusoidal chirp and a transient linear chirp (see Fig. \ref{fig:Sinusoidal}). As with the previous example, here \textbf{Method E} seems to be superior for a high level of noise (SNR of 5 dB), while \textbf{Method A} offers the best results for a moderate level of noise (SNR of 20 dB). For this example, we can see the value of the methods based on SST2. While offering poorer results for moderate to high levels of noise, they are the only ones still performing at low levels of noise (SNR higher than 30 dB). This is because SST2 produces a good separation of the modes (see bottom left panel on Fig. \ref{fig:Sinusoidal}), while for the STFT they are too close so that the masking methods cannot produce two different masks.

\subsection{Real Signals}
\subsubsection{Bat echolocation call} 
\begin{center}
\begin{figure}
\includegraphics[width=\columnwidth]{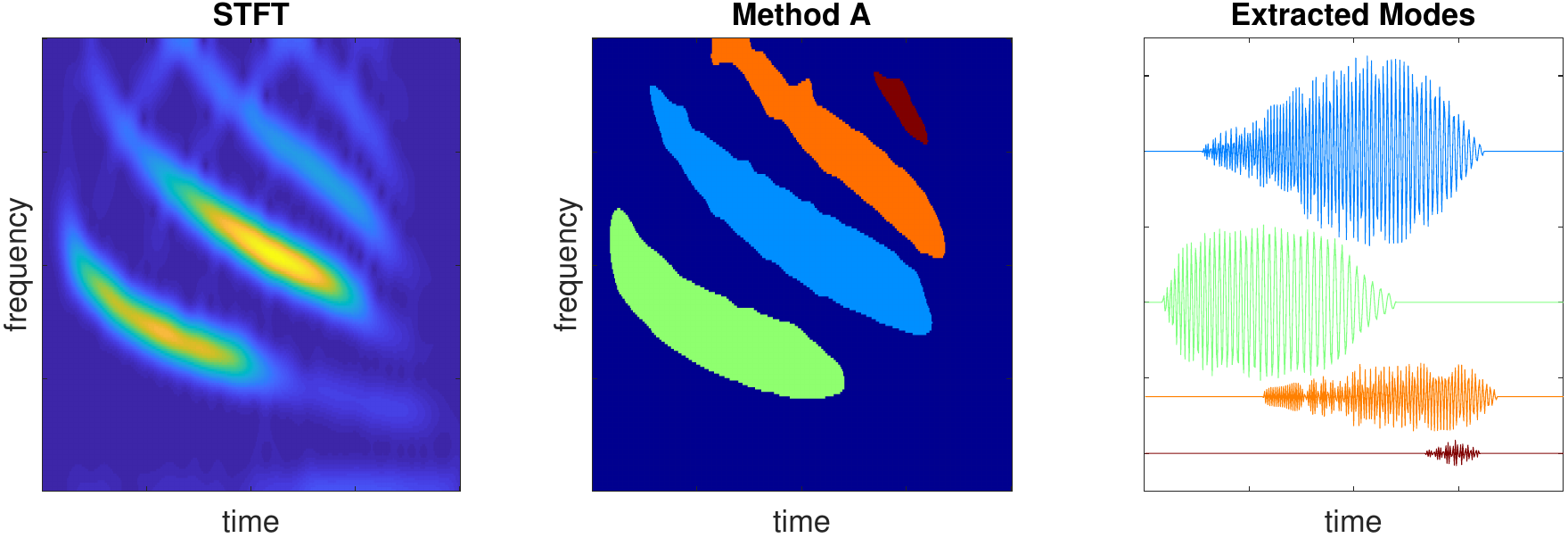}
    \caption{Bat signal. Left: modulus of the STFT. Middle: selected masks. Right: extracted modes (shifted vertically for visibility purposes).}\label{fig:BAT}
    \end{figure}
\end{center}

We start by showing a typical example for TF filtering: the classical bat signal. The results are presented on Fig. \ref{fig:BAT}. We can see that \textbf{Method A} produces a satisfactory result isolating four components. It is important to mention that some methods fail at isolating the short high-frequency mode (plotted in maroon color).
\subsubsection{Fetal ECG extraction} 
\begin{center}
\begin{figure}
    \includegraphics[width=\columnwidth]{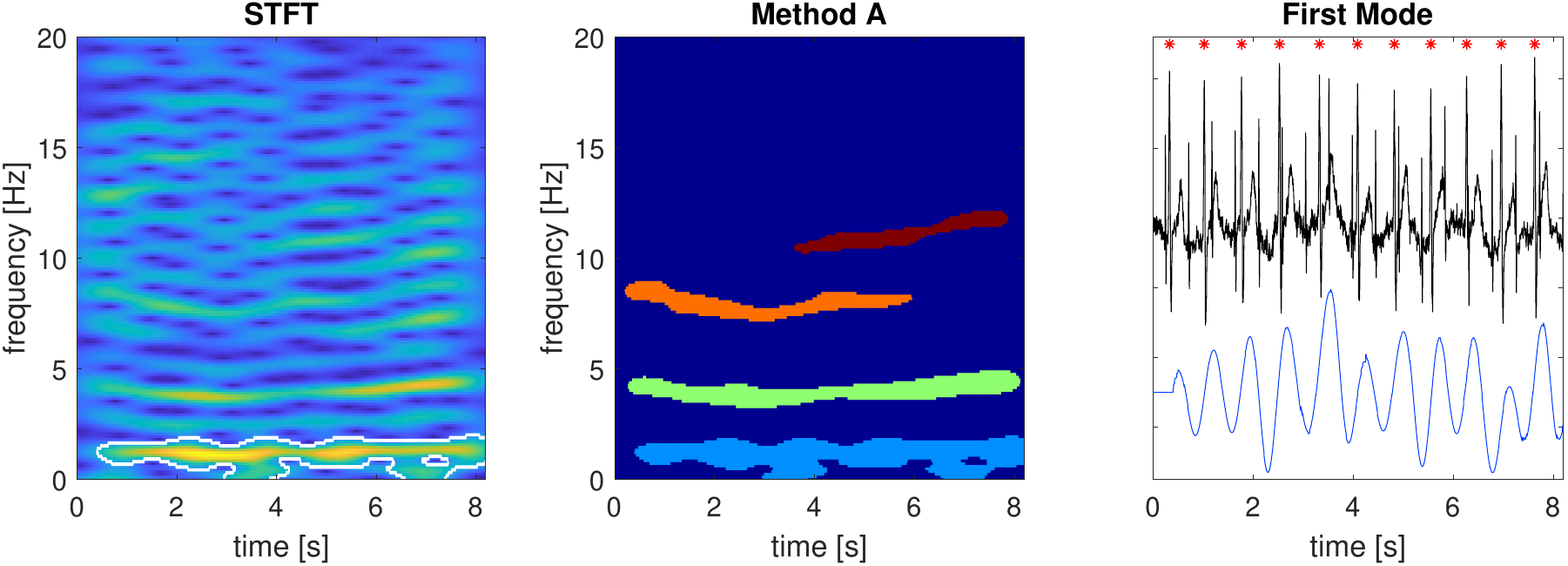}
    \caption{Fetal ECG signal. Left: modulus of the STFT, with the contour of the mask of the first (fundamental) mode. Middle: selected masks. Right: input signal (in black), first (fundamental) mode (in blue), and R-peaks of the maternal ECG (red asterisks `{\color{red} *}' ).}\label{fig:ECG_FETAL}
    \end{figure}
\end{center}

A second real example can be found on Fig. \ref{fig:ECG_FETAL}. Here we analyzed an 8-second segment of an ECG recording where there is a maternal component, and a fetal one. For this case, both independent signals create ribbons on the TF plane that crosses each other, making the separation a difficult problem. We can see that \textbf{Method A} success in estimating meaningful TF masks. The one in blue color estimates the fundamental component of the maternal signal. This can be confirmed on the right panel of Fig. \ref{fig:ECG_FETAL}. There, we show the input signal (in black), and the fundamental mode (in blue). The maternal R-peaks are also marked with red asterisks. We can observe 11 maternal cycles, its fundamental frequency being encoded in the blue mode. A posterior analysis with wave-shape functions would make use of this fundamental component for amplitude and phase estimation.

\section{Conclusions}\label{sec:Conclusions}
We presented here a versatile and flexible family of methods that can be used for TF filtering. Using a graph theory framework and relying on the strength of work done in the area of graph clustering, we were able to formalize several algorithms for masking the TF plane. An inversion of these masks retrieves the desired components.  Advantages of the method include robustness and stability; the method is non-parametric and does not require any underlying assumptions on the signal. Finally, the method is one-shot and, in contrast to sequential methods that remove one component at a time, works globally and avoids an accumulation of errors.
We illustrated the capabilities of our proposal both in simulated and real examples. In all cases, the results show a satisfactory performance, evidencing a method that is able to identify meaningful components.
We have seen that five different algorithms that fit into this family perform reasonably well. It is a natural avenue for future work to find out whether there is a particularly distinguished algorithm in this family; it is conceivable that a particular choice of ingredients leads to additional synergestic effects that can boost the performance (as a purely hypothetical example, suppose we connected pixels whenever their $\ell^{3/2}-$norm is small and use $ \min \left\{  |A_{ij}|,|A_{k \ell}|  \right\} \geq \tau$, then maybe that combination is particularly useful when combined with spectral clustering with the Kirchhoff-Laplacian).

The graph theory approach can be straightforwardly extended to other TF representations such as the continuous wavelet transform, Cohen's or affine classes or de-shape algorithm. This will be explored in the future.

\bibliographystyle{IEEEtran}
\bibliography{references}

\end{document}